# High-resolution magnetic imaging by mapping the locally induced anomalous Nernst effect using atomic force microscopy

Nico Budai[1], Hironari Isshiki[1,2]*, Ryota Uesugi[1,2], Zheng Zhu[1], Tomoya Higo[1,2,3], Satoru Nakatsuji[1,2,3,4] and YoshiChika Otani[1,2,4,5]

[1]Institute for Solid State Physics, The University of Tokyo; Kashiwa, Chiba, 277-8581, Japan.
[2]CREST, Japan Science and Technology Agency (JST); Saitama, 332-0012, Japan.
[3]Department of Physics, The University of Tokyo; Bunkyo-ku, Tokyo, 113-0033, Japan.
[4]Trans-scale Quantum Science Institute, The University of Tokyo; Bunkyo-ku, Tokyo, 113-0033, Japan.
[5]Center for Emergent Matter Science RIKEN; Wako, Saitama, 51-0198, Japan.
*Corresponding author. Email: h_isshiki@issp.u-tokyo.ac.jp

**Abstract**
We report a magnetic imaging method using atomic force microscopy to measure the locally induced anomalous Nernst effect. The tip contact creates a local temperature gradient on the sample surface controlled by a neighboring Joule heating wire. We demonstrate the imaging of magnetic domains in a nanowire of the ferromagnetic Weyl semimetal $Co_2MnGa$ with a spatial resolution of a sub-hundred nanometer at room temperature.

Magnetic Weyl semimetals[1,2] have attracted wide attention since they exhibit giant non-diagonal responses such as the anomalous Nernst effect[3,4] (ANE) and the anomalous Hall effect[5,6], originating from the characteristic band structures. Recently, the ferromagnetic Weyl semimetal $Co_2MnGa$ (CMG) was found[3] to exhibit a giant ANE, one order of magnitude greater than previously reported values. The anomalous Nernst thermopile, which converts thermal to electric energy, has been proposed for efficient energy harvesting[7,8]. The measurement of the local ANE in the thermopile is vital to optimize the device structures. Also, the antiferromagnetic Weyl semimetals $Mn_3X$ (X = Sn, Ga) are promising candidates for the ingredients in antiferromagnetic memory and logic devices. The magnetization reversal[9] and magnetic domain wall motion[10] induced by spin-orbit and spin-transfer torques have been recently investigated to realize these devices. Magnetic imaging is essential to evaluate these torques. The magneto-optical Kerr effect[11,12] and the scanning anomalous Nernst effect microscopy with laser[13–15] offer handy magnetic imaging methods. However, the spatial resolution of these optical techniques is typically sub-micrometers and, at most several hundred nanometers. Therefore, alternative magnetic imaging methods have been awaited to clarify the potential properties of the magnetic Weyl semimetals.

Here, we report a magnetic imaging method with high spatial resolution by mapping the local ANE employing atomic force microscopy (AFM). This paper demonstrates magnetic imaging on the Weyl ferromagnet CMG that has a significant ANE; $S_{ANE}$ ~ −5 μV/K [3,16,17]. Importantly, this method applies to the antiferromagnetic Weyl semimetals. Our approach does not need the expensive cantilever with an integrated heater[18] used for conventional scanning thermal microscopes[19–22]. Notably, the local ANE detection by the conventional scanning thermal microscopies has not been reported.

Figure 1 (a) illustrates the concept of our method. The Joule effect of the heater causes a non-equilibrium temperature distribution in the sample. The ANE is expressed as, $\mathbf{E}_{ANE} = S_{ANE} \cdot (\mathbf{m} \times \nabla T)$, where $\mathbf{E}_{ANE}$ is the ANE-induced electric field, $\mathbf{m}$ is the unit vector of the magnetization, and $\nabla T$ is the temperature gradient. The AFM tip contact with the sample surface locally generates a vertical temperature gradient $\nabla T_z$. If the magnetization under the tip points to the wire width direction (the *y*-direction), the ANE causes a voltage along the *x*-direction. Thus, the tip-induced ANE voltage represents the *y*-component of the local magnetization. We achieve magnetic imaging by mapping the local ANE voltage by scanning the tip on the sample surface.

We employ a dc magnetron sputtering method for growing CMG films on MgO(100) substrates at room temperature as described in Ref. 23. We used a mosaic target method with Co–Mn alloy tips on Co–Mn–Ga alloy sputtering target[24] to obtain stoichiometric composition of $Co_2MnGa$; the composition of the CMG was Co:Mn:Ga = 1.98:1.01:1.01. We performed post-annealing at 550 °C for 30 min and deposited 5-nm-thick $Al_2O_3$ on the CMG film as a capping layer. Two parallel CMG nanowires, which are 1-*μ*m-separated from each other, with a height (width) of 80 nm (600 nm), were fabricated by the standard technique of electron beam lithography and Ar ion etching. The capping layer was removed by subsequent Ar ion etching after removing the resist. The device structure is shown in Fig. 1 (b).

We used an atomic force microscope CoreAFM from Nanosurf[25], operable under the variable magnetic field generator[26]. The maximum magnetic field strength was 134 mT in our setup. We adopted the silicon nitride ($Si_3N_4$) cantilever with a spring constant of 0.48 N m$^{-1}$ (PNP-DB, NanoWorld AG, Switzerland). Silicon nitride cantilever is adequate here, since it is harder than the metallic sample and has a high thermal conductivity among insulators. Our device was placed on a home-made sample holder with a terminal for electrodes. In the measurement, we applied an AC current (typically, $I_{AC}$ = 4 mA with the frequency $f$ = 1043.43 Hz) to the heater. The second harmonic voltage $V_{2\omega}$ across the sample was detected using the standard lock-in technique with a time constant of 20 ms through a signal I/O option(Nanosurf, Liestal). All measurements were performed in atmosphere at room temperature.

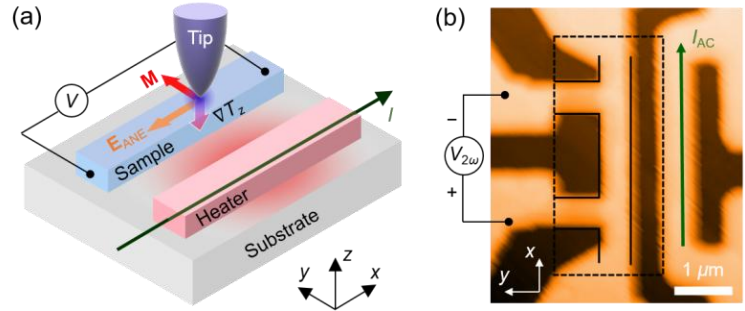

**Figure 1. Conceptual drawing of the magnetic imaging method and the sample structure.** (a) The illustration of the magnetic imaging method by mapping the local anomalous Nernst effect induced by the tip contact. The symbols *I*, *V*, **M**, $\nabla T_z$ and $\mathbf{E}_{ANE}$ represent electric current, observed voltage, magnetization, vertical temperature gradient and the electric field induced by the anomalous Nernst effect, respectively. (b) A topographic image of a typical sample device taken by atomic force microscopy, on which the electrical probe configuration for the measurement is indicated.

We apply an electric current ($I_{AC}$ = 4 mA) to the heater on the right and detect the second harmonic voltage ($V_{2\omega}$) across the sample on the left, as shown in Fig. 1 (b), under an external magnetic field of +134 mT in y-direction. The tip is placed above the sample as marked by a blue cross in the inset of Fig. 2 (a), where the tip-surface distance is initially ~ 3 μm. We make the tip approach until it contacts the sample surface with a loading force of $F_L = 15$ nN and retract it after keeping the contact condition for 10 s. During this process, we measure the atomic force $F$ and the voltage $V_{2\omega}$ as a function of time. The typical result is shown in Fig. 2 (a). The variation of the atomic force seen in the upper window of Fig. 2 (a) represents the processes: (1) approach, (2) contact, (3) keep the contact, (4) "pull-off", and (5) retract. The force overshooting to the negative direction in the region (4) is caused by the tip sticking to the surface. In the lower window in Fig. 2 (a), we can see that the voltage $V_{2\omega}$ changes once the tip contacts the sample surface in the region (2) and goes back to its initial value when the tip is completely retracted in the region (4). The gradual increase and decrease of the $V_{2\omega}$ while the tip is close to the sample in regions (1) and (5) are caused by a surrounding air gap, water meniscus conduction and the radiation[18,27]. We define the difference in $V_{2\omega}$ as $\Delta V_{2\omega} = V_{2\omega}^C - V_{2\omega}^0$, where $V_{2\omega}^C$ and $V_{2\omega}^0$ represent the voltage with and without tip contact, respectively.

The same procedure is repeated by applying different external magnetic fields along the y-direction. We show the value of $\Delta V_{2\omega}$ as a function of the magnetic field in Fig. 2 (b). The sign of $\Delta V_{2\omega}$ changes when switching the magnetic field, assuring the signal is magnetic. Since we have excluded the linear response to the current such as the anomalous Hall effect, due to the second harmonic detection, the signal is most probably attributable to the ANE induced by the local temperature gradient. The obtained $\Delta V_{2\omega}$ reflects the y-component of the magnetization of the CMG wire since we are measuring the ANE voltage along the x-direction and the induced temperature gradient is along the z-direction. The magnetic field applied orthogonally to the wire, lower than 134 mT, is not strong enough to saturate the magnetization. This explains the almost linear $\Delta V_{2\omega}$ to the magnetic field in Fig. 2 (b). We show in Fig. 2 (c) the controlled loading force $F_L$ dependence of the $\Delta V_{2\omega}$ under +134 mT along the y-direction. We confirm the force of about 12.5 nN is enough to saturate the signal. The signal $\Delta V_{2\omega}$ was stable during the whole experiments carried out in the present work.

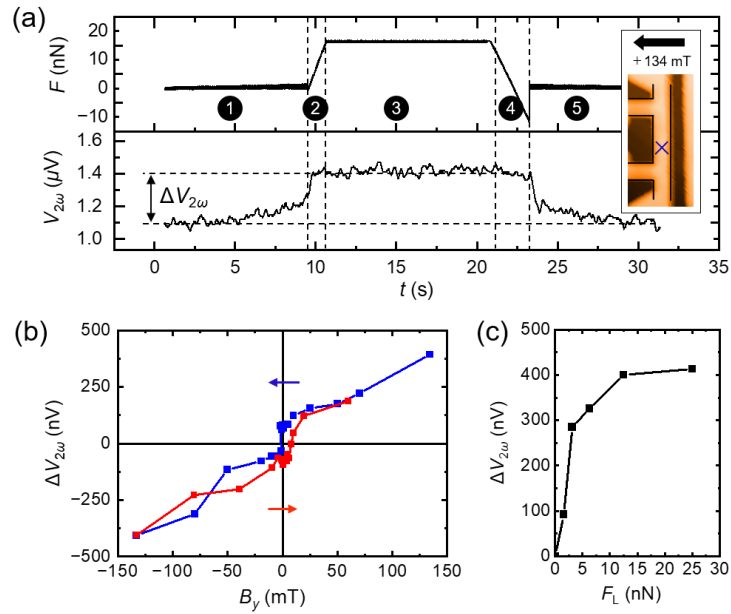

**Figure 2. Modulation of the second harmonic voltage upon the tip contact.** (a) The time evolution of atomic force $F$ and the voltage $V_{2\omega}$ during the tip approaching, contacting, and retracting to/from the sample. The inset shows the tip position and the external magnetic field direction. We define the magnitude of the signal induced by the tip contact as $\Delta V_{2\omega}$. (b) The signal $\Delta V_{2\omega}$ as a function of the magnetic field along the y-direction. The red and blue arrows indicate the sweeping direction of the magnetic field. (c) The signal $\Delta V_{2\omega}$ as a function of the controlled loading force $F_L$.

We map the locally induced ANE on the sample by detecting $V_{2\omega}$ at each point through scanning the sample surface in static force (contact) mode. This is possible since the response of the ANE voltage is quick enough by tip contact, as shown in Fig. 2 (a). The scanning is done along the y-direction in the area as marked by the dashed lines in Fig. 1 (b) with the loading force $F_L = 20$ nN. The typical scanning rate is ~ 0.2 Hz to obtain an image with 100 × 100 pixels. The results under the magnetic fields of +134, −5, −17, and −134 mT along the y-direction are shown in Figs. 3 (a), (b), (c), and (d), respectively. Here, the color represents the value of $\Delta V_{2\omega}$ as shown in Fig. 3(e), which reflects the y-component of the magnetization. The color on the sample in Fig. 3 (a) is almost white, indicating the magnetization nearly aligns with the magnetic field. The color drastically changes when we sweep the magnetic field to −5 mT (Fig. 3(b)). However, a white spot remains in the middle, indicating that the local magnetization at this spot is still pointing in the +y-direction. This is consistent with the small hysteresis loop in Fig. 2 (b). A pair of white and black spots in the middle of the wire at −17 mT (Fig. 3 (c)) implies the presence of a magnetic domain wall. The color becomes almost black when the magnetic field is −134 mT (Fig. 3 (d)), showing magnetization reversal. It is reasonable to see no color contrast on the wires of the voltage electrodes. In these electrodes, the magnetization aligns to the y-direction due to the shape anisotropy, that does not create detectable voltage in this geometry.

Using the same method, we attempted magnetic imaging on a conventional ferromagnet Permalloy (Py: $Ni_{80}Fe_{20}$). See supplementary material for the details. We saw a change of $\Delta V_{2\omega}$ on the Py nanowire depending on the magnetic field direction which is similar to the result on CMG. Different values of the size of the anomalous Nernst effect for Py have been reported[28–30], but according to the conventional scaling relation with the magnetization $S_{ANE}$ should be smaller than ~ 0.7 $\mu$V/K[31]. Therefore, our method applies to the antiferromagnetic Weyl semimetal $Mn_3Sn$ ($S_{ANE}$ ~ 0.5 $\mu$V/K[31]).

We simulated the magnetic domain structures under each magnetic field by OOMMF[32]. We used the exact dimensions as the actual sample for the simulation. The cell size in the model is 10 nm, which is short enough compared to the width of the wire. The exchange stiffness of CMG was set to be $A = 4 \cdot 10^{-11}$ J/m[33], and the local saturation magnetization to be $M_S = 760$ kA/m[34]. The results of the simulation in Figs. 3 (f)-(i) correspond to the experimental results in Fig. 3 (a)-(d), respectively. A white and black color represents that the magnetization aligns to the +y-direction and −y-direction, respectively, while grey lies between those two. The simulation reproduces similar features observed in the experimental results. The domain walls in Figs. 3 (c) appear in Fig. 3 (h). The inhomogeneous color contrast near the edge of the wire also appears in Figs. 3 (f) and (i), similar to the experimental results in Figs. 3 (a) and (d), respectively. We estimate the spatial resolution of our magnetic imaging method by checking the "edge response" of $\Delta V_{2\omega}$. We show, in Fig. 3 (j), the line profile of $\Delta V_{2\omega}$ along the green line in Fig. 3 (a) that crosses the edge of the sample. The signal of $\Delta V_{2\omega}$ should sharply change when the tip steps across the edge of the sample. The distance required for the rise of the signal ($\Delta V_{2\omega}$) from 10% to 90% is a good parameter for the spatial resolution[35], that is ~ 80 nm in our method.

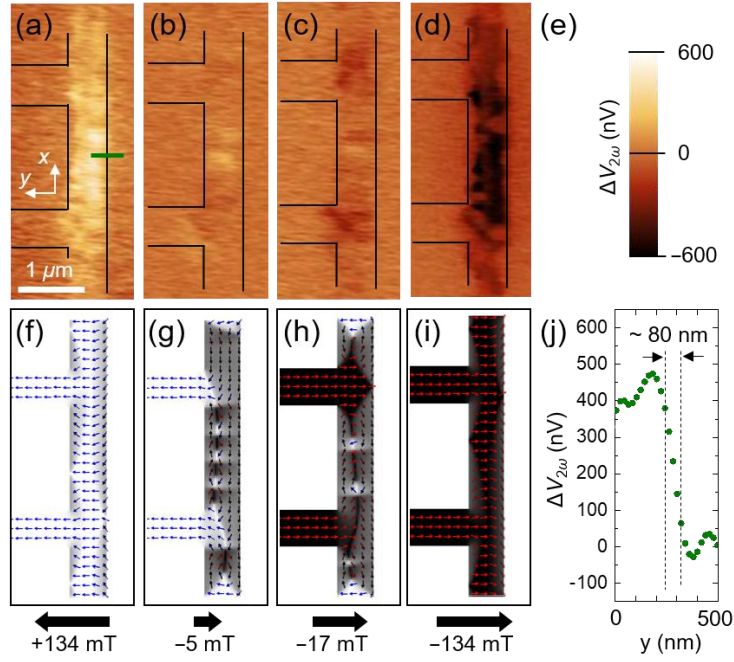

**Figure 3. Magnetic imaging of the Co$_2$MnGa nanowire and magnetic domain structures simulated by OOMMF.** (a)-(d) The $\Delta V_{2\omega}$ mappings on the Co$_2$MnGa nanowire surface with the external magnetic fields of +134, −5, −17, and −134 mT, respectively, along the *y*-direction. The measurement configuration is shown in Fig. 1 (b). (e) The color scale of $\Delta V_{2\omega}$ for (a)-(d). (f)-(g) The simulated magnetic domain structures by OOMMF for applied magnetic fields of 134, −5, −17, and −134 mT, respectively. The small arrows indicate the local magnetization direction, while the color in the background (white-grey-black) indicates the *y*-components of the local magnetization. (j) The cross section in the green line in (a).

As shown in Fig. 4, we simulate the temperature distribution in the device using COMSOL Multiphysics[36]. The geometry of the sample is the same as the actual device, while the tip is represented by a truncated cone. The temperatures of the substrate boundary and the bottom base of the truncated cone (the root of the tip) are fixed at 286 K. The top surface of the cone (the tip apex) directly touches the sample surface with a contact area[37] of $4\pi$ nm$^2$ in our model. The material parameters used in the simulation are summarized in Table I. The simulated temperature distribution is shown in Fig. 4 (a), where the applied current in the heater is 5 mA. The time variation of the temperature distribution follows the AC current quickly enough for the lock-in measurement when the frequency is < 10 kHz. The temperature of the heating wire reaches ~300 K, an increase of ~14 K from the initial value. In contrast, the temperature of the sample barely rises to ~289 K, an increase of ~ 3 K. A considerable vertical temperature gradient builds up in the vicinity of the sample surface by contacting the tip, as shown in the magnified view in Fig. 4 (a). The vertical temperature gradient is generated primarily under the contact area, resulting in a high spatial resolution of ~ 80 nm.

**Table I. Material parameters for the simulation.**

| Materials | $\rho$ [$\mu\Omega$cm] | $\kappa$ [W/mK] |
|---|---|---|
| Co$_2$MnGa | 120 | 15.93[3] |
| Si$_3$N$_4$ | - | ~60[38] |
| MgO | - | 55[39] |

The ANE voltage induced by tip contact $\Delta V_{2\omega}$ is given by,

$$\Delta V_{2\omega} = \frac{S_{ANE}}{t \cdot w} \iiint (\nabla T_z^C - \nabla T_z^0)\, dxdydz = S_{ANE} \cdot l \cdot (\overline{\nabla T_z^C} - \overline{\nabla T_z^0}), \quad (1)$$

where $t$, $w$, $l$, $\nabla T_z^C$, and $\nabla T_z^0$ are the sample thickness, width, length, and the vertical temperature gradient with and without tip contact, respectively. The integral is taken over the sample volume. The overline indicates the average of the temperature gradient over the sample volume. According to the simulation, the averaged vertical temperature gradient varies from $\overline{\nabla T_z^0} = -6.1 \times 10^4$ K/m to $\overline{\nabla T_z^C} = -9.7 \times 10^4$ K/m by tip contact. We evaluate the magnitude of the ANE ($S_{ANE}$) of the CMG wire from the experiment and the simulation. The heating power dependence of the experimental $\Delta V_{2\omega}$ under +134 mT and the simulated $\Delta(\overline{\nabla T_z}) (\equiv \overline{\nabla T_z^C} - \overline{\nabla T_z^0})$ are summarized in Fig. 4 (b) (upper panel). Both, $\Delta V_{2\omega}$ and $\Delta(\overline{\nabla T_z})$, are almost linear to the power. The calculated $S_{ANE}$ as a function of the applied power is shown in Fig. 4 (b) (lower panel). The average over all calculated values, $S_{ANE} \sim -4.7$ μV/K, is consistent with that of CMG from the literature[3,16,40,41]. As seen in the magnified view in Fig. 4 (a), the temperature gradient along the y-direction is huge ($\overline{\nabla T_y^0} = 3.3 \times 10^6$ K/m), while the induced gradient by tip contact is few ($\Delta(\overline{\nabla T_y}) \sim 10^2$ K/m). Therefore, the temperature gradient along the y-direction would contribute only to the offset voltage in Fig. 2 (a), if there was a z-component of the magnetization in the sample. The result of the simulation is consistent with the scenario we described above.

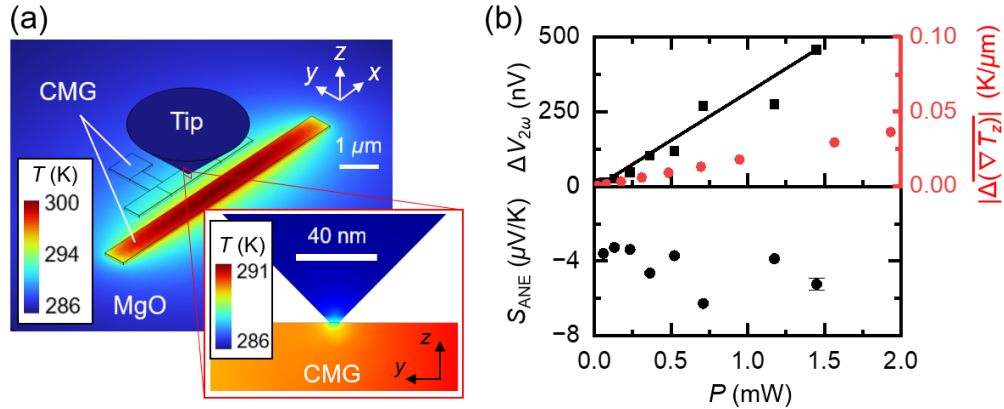

**Figure 4. Simulated temperature distribution in the sample.** (a) Temperature distribution over the whole device with an electric current of 5 mA applied to the heater wire simulated by COMSOL Multiphysics. The effect of the tip contact can be seen in the magnified view. (b) The heater power dependence of $\Delta V_{2\omega}$, $|\Delta(\overline{\nabla T_z})|$ and $S_{ANE}$. Here, $\Delta V_{2\omega}$ is obtained experimentally, while $|\Delta(\overline{\nabla T_z})|$ is simulated theoretically.

In this paper, we detected the locally induced anomalous Nernst effect by employing the tip of an atomic force microscope on the nanowire of the Weyl ferromagnet Co$_2$MnGa. We achieved magnetic imaging by mapping the local anomalous Nernst effect with a spatial resolution of ~ 80 nm. Our method requires no special cantilever with an integrated heater as for conventional scanning thermal microscopy. Moreover, unlike conventional magnetic force microscopy, our technique applies to antiferromagnetic Weyl semimetals that exhibit the anomalous Nernst effect.

## SUPPLEMENTARY MATERIAL
See supplementary material for the magnetic imaging of the Py nanowire.

## ACKNOWLEDGEMENTS
This work was partially supported by: CREST (Grant No.JPMJCR18T3) from JST, JSPS KAKENHI Grant Number 19K15431 and 19H05629.

# High-resolution magnetic imaging by mapping the locally induced anomalous Nernst effect using atomic force microscopy

Nico Budai[1], Hironari Isshiki[1,2*], Ryota Uesugi[1,2], Zheng Zhu[1], Tomoya Higo[1,2,3], Satoru Nakatsuji[1,2,3,4] and YoshiChika Otani[1,2,4,5]

[1]Institute for Solid State Physics, The University of Tokyo; Kashiwa, Chiba, 277-8581, Japan.
[2]CREST, Japan Science and Technology Agency (JST); Saitama, 332-0012, Japan.
[3]Department of Physics, The University of Tokyo; Bunkyo-ku, Tokyo, 113-0033, Japan.
[4]Trans-scale Quantum Science Institute, The University of Tokyo; Bunkyo-ku, Tokyo, 113-0033, Japan.
[5]Center for Emergent Matter Science RIKEN; Wako, Saitama, 51-0198, Japan.
* h_isshiki@issp.u-tokyo.ac.jp

**Magnetic imaging of a permalloy ($Ni_{80}Fe_{20}$) nanowire**

Here, we show magnetic imaging on a permalloy (Py: $Ni_{80}Fe_{20}$) nanowire using the same method as described in the main text of the paper. Two Py nanowires with a thickness of 80 nm were fabricated by the standard technique of electron beam lithography and electron beam deposition. The widths of the sample and heater were 200 nm and 2 $\mu$m, respectively. We need a 2 $\mu$m wide heater to generate an adequate electric power, since the resistivity of Py (~ 40 $\mu\Omega\cdot$cm) is lower than that of $Co_2MnGa$. Copper (Cu) electrodes were deposited for the $V_{2\omega}$ detection. The device structure and the electrical probe configuration are shown in Fig. S1 (a). We applied an alternating current of 17 mA to the heater for the measurement below. The other parameters are described in the main text.

The results of the $\Delta V_{2\omega}$ mapping in the area of Fig. S1 (a) under the magnetic fields of +134 and -134 mT along the *y*-direction are shown in Figs. S1 (b) and (c), respectively. The positive (white) and negative (black) signals in the areas near the junction of the Cu electrode and Py sample are non-magnetic, since the signal does not change by switching the magnetic field direction. The tip contact on the device decreases the temperature at the junction, which induces these signals due to the Seebeck effect. (This thermoelectric effect creates a slope along *x*-direction in the $\Delta V_{2\omega}$ mapping, that has been eliminated in Fig. S1(b) and (c).) On the other hand, one can see the magnetic signal on the sample (in the grey boxes), that changes its contrast, depends on the magnetic field direction. This is the ANE signal induced by the tip contact. Therefore, the method works on a conventional ferromagnet whose ANE is smaller than that of $Co_2MnGa$.

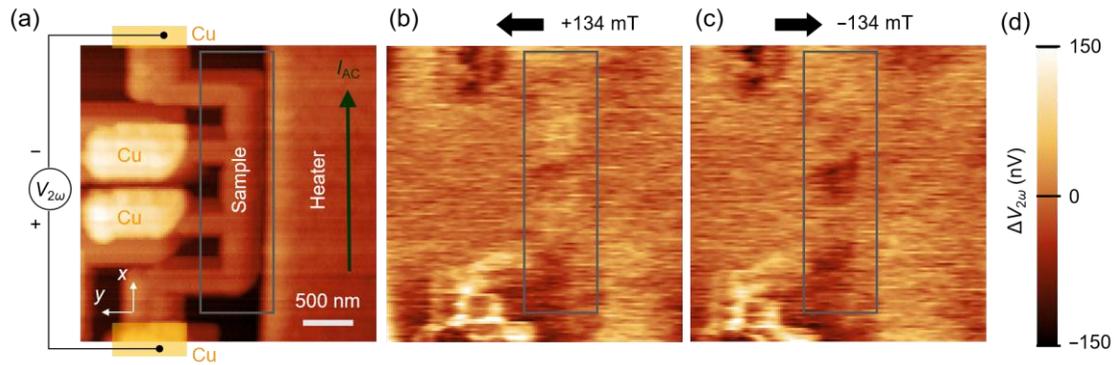

**Figure S1. Magnetic imaging of a permalloy (Py) nanowire.** (a) Topographic image of the Py device taken by atomic force microscopy, on which the electrical probe configuration for the measurement is indicated. Orange squares represent copper (Cu) electrodes for the $V_{2\omega}$ detection. (b), (c) The $\Delta V_{2\omega}$ mappings on Py with an external magnetic fields of +134 and -134 mT, respectively, along the *y*-direction. The slope in contrast along *x*-direction due to the Seebeck effect at Cu/Py interface has been eliminated. The grey boxes indicate the position of the sample. (d) The color scale of $\Delta V_{2\omega}$ for (b) and (c).